\documentclass[preprint,aps]{revtex4}
\usepackage{epsfig}
\usepackage{amsmath,amsfonts,amssymb,amsthm,bm}
\usepackage{graphicx,latexsym}
\usepackage{amsmath}
\begin{document}

\title{Dispersion Relations and Polarizations of Low-frequency Waves in
Two-fluid Plasmas}
\author{Jinsong~Zhao}
\affiliation{Purple Mountain Observatory, Chinese Academy of Sciences, Nanjing 210008,
China.}

\begin{abstract}
Analytical expressions for the dispersion relations and polarizations of
low-frequency waves in magnetized plasmas based on two-fluid model are
obtained. The properties of waves propagating at different angles (to the
ambient magnetic field $\mathbf{B}_{0}$) and $\beta $ (the ratio of the
plasma to magnetic pressures) values are investigated. It is shown that two
linearly polarized waves, namely the fast and Alfv\'{e}n modes in the low-$%
\beta $ $\left( \beta \ll 1\right) $ plasmas, the fast and slow modes in the
$\beta \sim 1$ plasmas, and the Alfv\'{e}n and slow modes in the high-$\beta
$ $\left( \beta \gg 1\right) $ plasmas, become circularly polarized at the
near-parallel (to $\mathbf{B}_{0}$) propagation. The negative
magnetic-helicity of the Alfv\'{e}n mode occurs only at small or moderate
angles in the low-$\beta $ plasmas, and the ion cross-helicity of the slow
mode is nearly the same as that of the Alfv\'{e}n mode in the high-$\beta $
plasmas. It also shown the electric polarization $\delta E_{z}/\delta E_{y}$%
\ decreases with the temperature ratio $T_{e}/T_{i}$\ for the
long-wavelength waves, and the transition between left- and right-hand
polarizations of the Alfv\'{e}n mode in $T_{e}/T_{i}\neq 0$\ plasmas can
disappear when $T_{e}/T_{i}=0$. The approximate dispersion relations in the
near-perpendicular propagation, low-$\beta $, and high-$\beta $ limits can
quite accurately describe the three modes.
\end{abstract}

\maketitle



\section{Introduction}

It is well known that in homogeneous magnetized two-fluid plasmas three
electromagnetic modes with frequency less than the electron cyclotron
frequency exist \cite{st63,fk69,sw89,is05,da09}. These include the fast,
Alfv\'en (or intermediate) and slow modes, according to their different phase
velocities \cite{kr94,be12}.
Their dispersion relations can be obtained from a general one based on the
Hall-MHD model \cite{st63,fk69,sw89,is05,da09}. Recently, \cite{be12}
obtained the same relation from a simpler formulation involving a
two-dimensional current density vector. The general dispersion relation for
even lower frequency modes (with wave frequency $\omega $ less than the ion
cyclotron frequency $\omega _{ci}$) have also been derived using different
formulations \cite{ho99,sh00,ch11}. However, a comprehensive investigation
of the wave polarizations is still lacking.

In this paper we present analytical expressions of the dispersion relations
and polarizations using an approach similar to that of Ref. \cite{ho99}. We
shall consider the dense-plasma limit $V_{A}^{2}/c^{2}\ll 1$, where $V_{A}$
is the Alfv\'{e}n speed and $c$ is the light speed, so that the displacement
current in the Ampere's law can be ignored \cite%
{st63,fk69,sw89,is05,da09,kr94,be12}. The resulting analytical expressions
are useful for analyzing the properties of low-frequency waves in different
plasmas.

In the next section we present the derivation of the dispersion relations
and polarizations of the waves. In Sec. III the properties of the waves
propagating at different angles and different $\beta $ regimes are
discussed. The main results are summarized in Sec. IV. The Appendix gives
the approximate dispersion relations in the near-perpendicular propagation,
low-$\beta $ $\left( \beta \ll 1\right) $, and high-$\beta $ $\left( \beta
\gg 1\right) $ limits, where $\beta $ is the ratio of the plasma to magnetic
pressures.

\section{Dispersion relations and polarizations}

Linearized two-fluid and Maxwell's equations are
\begin{eqnarray}
m_{\alpha }n_{0}\partial _{t}\mathbf{\delta v}_{\alpha } &=&n_{0}q_{\alpha
}\left( \mathbf{\delta E}+\mathbf{\delta v}_{\alpha }\times \mathbf{B}%
_{0}\right) -\nabla \delta P_{\alpha },  \label{Eq:MOMENTUM} \\
\partial _{t}\delta n_{\alpha } &=&-\nabla \cdot \left( n_{0}\mathbf{\delta v%
}_{\alpha }\right) ,  \label{Eq:CONTINUUM} \\
\nabla \times \delta \mathbf{B} &=&\mu _{0}\delta \mathbf{J},
\label{Eq:AMPERE} \\
\nabla \times \delta \mathbf{E} &=&-\partial _{t}\delta \mathbf{B},
\label{Eq:FARADAY}
\end{eqnarray}%
where the subscripts $\alpha =i,e$ denote ions and electrons, respectively, $%
m_{\alpha }$ is the mass, $q_{\alpha }$ is the charge, $\delta P_{\alpha
}=\kappa \gamma _{\alpha }T_{\alpha }\delta n_{\alpha }$ is the thermal
pressure, $\kappa $\textbf{\ }is Boltzmann constant, $T_{\alpha }$ is the
temperature, $\delta n_{\alpha }$ is the perturbed number density, $\mathbf{%
\delta v}_{\alpha }$ is the perturbed velocity, $\mathbf{\delta J}$ is the
perturbed current density, $\delta \mathbf{E}$ and $\delta \mathbf{B}$ are
the perturbed electric and magnetic fields, respectively, $\mathbf{B}%
_{0}=B_{0}\hat{\mathbf{z}}$ is the ambient magnetic field, and $n_{0}$ is
the ambient number density. As mentioned, the displacement current is
neglected. The quasi-neutrality condition $\delta n_{i}=\delta n_{e}\equiv
\delta n$ shall also be used. In the study an electron-proton plasma is
considered, namely $q_{e}=-e$\ and $q_{i}=e$.

We shall consider plane waves, so that $\delta f\propto \delta f_{k}$exp$%
(-i\omega t+i\mathbf{k}\cdot \mathbf{r})$, where $\omega $ is the wave
frequency and $\mathbf{k\equiv }\left( k_{\perp }\hat{\mathbf{e}}_{x}+k_{z}%
\hat{\mathbf{e}}_{z}\right) $ is the wave vector. We can obtain from Eq. (%
\ref{Eq:MOMENTUM}) the perpendicular and parallel (to $B_{0}\hat{\mathbf{z}}$%
) fluid velocities
\begin{equation}
\left( 1-\frac{\omega ^{2}}{\omega _{c\alpha }^{2}}\right) \delta \mathbf{v}%
_{\alpha \perp }=\frac{1}{B_{0}}\delta \mathbf{E}_{\perp }\times \hat{%
\mathbf{z}}-i\frac{\omega }{B_{0}\omega _{c\alpha }}\delta \mathbf{E}_{\perp
}-i\frac{\kappa \gamma _{\alpha }T_{\alpha }}{m_{\alpha }\omega _{c\alpha }}%
\mathbf{k}_{\perp }\times \hat{\mathbf{z}}\frac{\delta n}{n_{0}}-\frac{%
\kappa \gamma _{\alpha }T_{\alpha }\omega }{m_{\alpha }\omega _{c\alpha }^{2}%
}\mathbf{k}_{\perp }\frac{\delta n}{n_{0}},  \label{Eq:Vel-per}
\end{equation}%
and
\begin{equation}
\delta v_{\alpha z}=i\frac{q_{\alpha }}{m_{\alpha }\omega }\delta E_{z}+%
\frac{\kappa \gamma _{\alpha }T_{\alpha }}{m_{\alpha }\omega }k_{z}\frac{%
\delta n}{n_{0}}.  \label{Eq:Vel-par}
\end{equation}%
The current density $\delta \mathbf{J}=n_{0}e\left( \delta \mathbf{v}%
_{i}-\delta \mathbf{v}_{e}\right) $ can then be expressed as
\begin{eqnarray}
\Lambda _{0}\Lambda _{2}\delta \mathbf{J}_{\perp } &=&\frac{n_{0}e}{B_{0}}%
\left( \Lambda _{2}-\Lambda _{0}\right) \delta \mathbf{E}_{\perp }\times
\hat{\mathbf{z}}-i\frac{n_{0}e\omega }{B_{0}\omega _{ci}}\left( \Lambda
_{2}+Q\Lambda _{0}\right) \delta \mathbf{E}_{\perp }  \notag \\
&&-i\frac{n_{0}e\kappa T_{t}}{m_{i}\omega _{ci}}\left( \Lambda _{2}%
\widetilde{T}_{i}+\Lambda _{0}\widetilde{T}_{e}\right) \mathbf{k}_{\perp
}\times \hat{\mathbf{z}}\frac{\delta n}{n_{0}}  \notag \\
&&-\frac{n_{0}e\kappa T_{t}\omega }{m_{i}\omega _{ci}^{2}}\left( \Lambda _{2}%
\widetilde{T}_{i}-Q\Lambda _{0}\widetilde{T}_{e}\right) \mathbf{k}_{\perp }%
\frac{\delta n}{n_{0}},  \label{Eq:cur-1}
\end{eqnarray}%
and
\begin{equation}
\delta J_{z}=\frac{in_{0}e^{2}}{m_{e}\omega }\left( 1+Q\right) \delta E_{z}+%
\frac{n_{0}e\kappa T_{t}}{m_{e}\omega }k_{z}\left( Q\widetilde{T}_{i}-%
\widetilde{T}_{e}\right) \frac{\delta n}{n_{0}},  \label{Eq:cur-2}
\end{equation}%
where $Q\equiv m_{e}/m_{i}$, $\Lambda _{0}\equiv 1-\omega ^{2}/\omega
_{ci}^{2}$, $\Lambda _{2}\equiv 1-Q^{2}\omega ^{2}/\omega _{ci}^{2}$, $%
T_{t}\equiv \gamma _{i}T_{i}+\gamma _{e}T_{e}$, $\widetilde{T}_{i}\equiv
\gamma _{i}T_{i}/T_{t}$ and $\widetilde{T}_{e}\equiv \gamma _{e}T_{e}/T_{t}.$
Combining Eqs. (3) and (4) leads to
\begin{equation}
\delta \mathbf{J}=-i\frac{k^{2}}{\mu _{0}\omega }\delta \mathbf{E}+i\frac{1}{%
\mu _{0}\omega }\mathbf{k}\mathbf{k}\cdot \delta \mathbf{E}.
\label{Eq:cur-3}
\end{equation}%
From Eqs.(\ref{Eq:cur-1}) -- (\ref{Eq:cur-3}), we get for the electric field
and number density perturbation,
\begin{eqnarray}
i\left[ \Lambda _{0}\Lambda _{2}V_{A}^{2}k_{z}^{2}-\left( 1+Q\right) \Lambda
_{1}\omega ^{2}\right] \delta E_{x}+\left( 1-Q^{2}\right) \frac{\omega ^{3}}{%
\omega _{ci}}\delta E_{y}-i\Lambda _{0}\Lambda _{2}V_{A}^{2}k_{\perp
}k_{z}\delta E_{z}  \notag \\
-\frac{\kappa T_{t}}{e}\left( \Lambda _{2}\widetilde{T}_{i}-Q\Lambda _{0}%
\widetilde{T}_{e}\right) \omega ^{2}k_{\perp }\frac{\delta n}{n_{0}} =0,
\label{Eq:Relation_x} \\
\left( 1-Q^{2}\right) \frac{\omega ^{3}}{\omega _{ci}}\delta E_{x}-i\left[
\Lambda _{0}\Lambda _{2}V_{A}^{2}k^{2}-\left( 1+Q\right) \Lambda _{1}\omega
^{2}\right] \delta E_{y}  \notag \\
-i\frac{\kappa T_{t}}{e}\left( \Lambda _{2}\widetilde{T}_{i}+\Lambda _{0}%
\widetilde{T}_{e}\right) \omega \omega _{ci}k_{\perp }\frac{\delta n}{n_{0}}
=0,  \label{Eq:Relation_y} \\
i\lambda _{e}^{2}k_{\perp }k_{z}\delta E_{x}-i\left( 1+Q+\lambda
_{e}^{2}k_{\perp }^{2}\right) \delta E_{z}-\frac{\kappa T_{t}}{e}\left( Q%
\widetilde{T}_{i}-\widetilde{T}_{e}\right) k_{z}\frac{\delta n}{n_{0}} =0,
\label{Eq:Relation_z}
\end{eqnarray}%
so that three electric field components can be written as
\begin{eqnarray}
\Pi \delta E_{x} &=&ik_{\perp }\frac{\kappa T_{t}}{e}\Pi _{\mathrm{Ex}}\frac{%
\delta n}{n_{0}},~~  \notag \\
\Pi \delta E_{y} &=&k_{\perp }\frac{\kappa T_{t}}{e}\Pi _{\mathrm{Ey}}\frac{%
\delta n}{n_{0}},~~  \notag \\
\Pi \delta E_{z} &=&ik_{z}\frac{\kappa T_{t}}{e}\Pi _{\mathrm{Ez}}\frac{%
\delta n}{n_{0}},  \label{Electric-field}
\end{eqnarray}%
where $\Lambda _{1}\equiv 1-Q\omega ^{2}/\omega _{ci}^{2}$ and other
definitions are:
\begin{eqnarray}
\Pi &=&\left( 1+Q\right) ^{2}\left( 1+Q+\lambda _{e}^{2}k_{\perp
}^{2}\right) \omega ^{4}  \notag \\
&-&\left( 1+Q\right) \left[ 1+Q+\lambda _{e}^{2}k_{\perp }^{2}+\left(
1+Q\right) k_{z}^{2}/k^{2}\right] \Lambda _{1}V_{A}^{2}k^{2}\omega ^{2}
\notag \\
&+&\left( 1+Q\right) \Lambda _{0}\Lambda _{2}V_{A}^{4}k^{2}k_{z}^{2},  \notag
\\
\Pi _{\mathrm{Ex}} &=&\left( 1+Q\right) \left( 1+Q+\lambda _{e}^{2}k_{\perp
}^{2}\right) \left( Q\widetilde{T}_{i}-\widetilde{T}_{e}\right) \omega ^{4}
\notag \\
&-&\left[
\begin{array}{c}
\left( 1+Q+\lambda _{e}^{2}k_{\perp }^{2}\right) \left( \Lambda _{2}%
\widetilde{T}_{i}-Q\Lambda _{0}\widetilde{T}_{e}\right) \\
+\left( 1+Q\right) \left( Q\widetilde{T}_{i}-\widetilde{T}_{e}\right)
\Lambda _{1}k_{z}^{2}/k^{2}%
\end{array}%
\right] V_{A}^{2}k^{2}\omega ^{2}  \notag \\
&+&\Lambda _{0}\Lambda _{2}\left( Q\widetilde{T}_{i}-\widetilde{T}%
_{e}\right) V_{A}^{4}k^{2}k_{z}^{2},  \notag \\
\Pi _{\mathrm{Ey}} &=&\left( 1+Q\right) \left[ \left( 1+Q+\lambda
_{e}^{2}k_{\perp }^{2}\right) \omega ^{2}-\Lambda _{1}V_{A}^{2}k_{z}^{2}%
\right] \omega \omega _{ci},  \notag \\
\Pi _{\mathrm{Ez}} &=&\Pi _{\mathrm{Ex}}+\left( 1-Q^{2}\right)
V_{A}^{2}k^{2}\omega ^{2}.  \notag
\end{eqnarray}

Inserting above electric field components into the number density equation
that is derived from Eqs. (\ref{Eq:CONTINUUM}), (\ref{Eq:Vel-per}) and (\ref%
{Eq:Vel-par}),
\begin{equation}
\left[ \left( \Lambda _{0}+\rho _{i}^{2}k_{\perp }^{2}\right) \omega
^{2}-\Lambda _{0}V_{Ti}^{2}k_{z}^{2}\right] \frac{\delta n}{n_{0}}=-i\frac{%
\omega ^{2}k_{\perp }}{B_{0}\omega _{ci}}\delta E_{x}+\frac{\omega k_{\perp }%
}{B_{0}}\delta E_{y}+ik_{z}\frac{e}{m_{i}}\Lambda _{0}\delta E_{z},
\label{Eq:ION-EQUATION}
\end{equation}%
the general dispersion relation can be expressed as
\begin{equation}
A\omega ^{6}-B\omega ^{4}+C\omega ^{2}-D=0,
\label{Eq:General dispersion equation}
\end{equation}%
with
\begin{eqnarray}
A &=&\left( 1+Q\right) \left( 1+Q+\lambda _{e}^{2}k^{2}\right) ^{2},  \notag
\\
B &=&\left[ \left( 1+Q\right) \left( 1+Q+\lambda _{e}^{2}k^{2}\right)
+\left( 1+Q+\lambda _{e}^{2}k^{2}\right) ^{2}\beta +\left( 1+Q^{3}\right)
\lambda _{i}^{2}k_{z}^{2}+\left( 1+Q\right) ^{2}k_{z}^{2}/k^{2}\right]
V_{A}^{2}k^{2},  \notag \\
C &=&\left[ \left( 1+Q\right) \left( 1+2\beta \right) +\left( 1+Q^{2}\right)
\rho ^{2}k^{2}\right] V_{A}^{4}k^{2}k_{z}^{2},  \notag \\
D &=&\beta V_{A}^{6}k^{2}k_{z}^{4}.  \notag
\end{eqnarray}%
where $\rho ^{2}=\rho _{i}^{2}+\rho _{s}^{2}$, $\rho _{i}\equiv
V_{Ti}/\omega _{ci}$\ is the ion gyroradius, $\rho _{s}\equiv V_{Ts}/\omega
_{ci}$\ is the ion acoustic gyroradius, $\lambda _{i}$ is the ion inertial
length, $\lambda _{e}$ is the electron inertial length,\ $V_{Ti}=\sqrt{%
\kappa \gamma _{i}T_{i}/m_{i}}$\ is the ion thermal speed, $V_{Ts}=\sqrt{%
\kappa \gamma _{e}T_{e}/m_{i}}$\ is the ion acoustic speed, $V_{T}=\sqrt{%
\kappa T_{t}/m_{i}}$\ is the sound speed, and $\beta \equiv
V_{T}^{2}/V_{A}^{2}$. With respect to the existing ones \cite%
{st63,fk69,be12,is05,da09,ho99,ch11}, Eq. (\ref{Eq:General dispersion
equation}) represents a more general description of the low-frequency
electromagnetic waves. Three roots for $\omega ^{2}$ correspond to the fast $%
\left( j=0\right) $, Alfv\'en $\left( j=1\right) $, and slow $\left(
j=2\right) $ modes, or \cite{be12,ch11},%
\begin{equation}
\omega _{j}^{2}=2p^{1/2}\mathrm{cos}\left( \frac{1}{3}\mathrm{cos}%
^{-1}\left( -\frac{q}{p^{3/2}}\right) -\frac{2\pi }{3}j\right) +\frac{B}{3A}%
,~~j=0,1,2,  \label{Eq:Roots}
\end{equation}%
with $p=\left( B^{2}-3AC\right) /(9A^2)$ and $q=\left(
9ABC-2B^{3}-27A^2D\right) /(54A^3)$. If we set $k\rightarrow \infty $, Eq. (%
\ref{Eq:General dispersion equation}) yields two resonances $\left(
ck/\omega \rightarrow \infty \right) $: the ion cyclotron resonance $\omega
=\omega _{ci}\cos \theta $\ and the electron cyclotron resonance $\omega
=\left\vert \omega _{ce}\right\vert \cos \theta $.

If we neglect the electron inertial terms $\left( \lambda _{e}k\right) $\
and terms of the order of $Q$, Eq. (\ref{Eq:General dispersion equation})
recovers the Hall-MHD dispersion relation \cite{bs03}, where only the ion
cyclotron resonance exists. For the high oblique propagation, low-$\beta $
and high-$\beta $\ limits, the approximate dispersion relations of the three
modes are given in the Appendix. Eq. (\ref{Eq:General dispersion equation})
can also be reduced to the well-known results in the cold two-fluid plasmas $%
\left( T_{i}=T_{e}=0\right) $\ \cite{stix92,ve13}.

Once the electric field perturbation (\ref{Electric-field}) and the
dispersion relation (\ref{Eq:Roots}) are known, the magnetic field and
velocity perturbations can be also expressed in terms of the number density
perturbation,
\begin{equation}
\frac{\delta B_{x}}{B_{0}}=-\rho ^{2}k_{\perp }k_{z}\frac{\omega _{ci}}{%
\omega }\frac{\Pi _{\mathrm{Ey}}}{\Pi }\frac{\delta n}{n_{0}},~~\frac{\delta
B_{y}}{B_{0}}=-i\rho ^{2}k_{\perp }k_{z}\frac{\left( 1-Q^{2}\right)
V_{A}^{2}k^{2}\omega \omega _{ci}}{\Pi }\frac{\delta n}{n_{0}},~~\frac{%
\delta B_{z}}{B_{0}}=\rho ^{2}k_{\perp }^{2}\frac{\omega _{ci}}{\omega }%
\frac{\Pi _{\mathrm{Ey}}}{\Pi }\frac{\delta n}{n_{0}},
\label{Magnetic-field}
\end{equation}%
\begin{equation}
\frac{\delta v_{ix}}{V_{T}}=\rho k_{\perp }\frac{\Pi _{\mathrm{vix}}}{\Pi }%
\frac{\delta n}{n_{0}},~~\frac{\delta v_{iy}}{V_{T}}=-i\rho k_{\perp }\frac{%
\Pi _{\mathrm{viy}}}{\Pi }\frac{\delta n}{n_{0}},~~\frac{\delta v_{iz}}{V_{T}%
}=\frac{V_{T}k_{z}}{\omega }\frac{\Pi _{\mathrm{viz}}}{\Pi }\frac{\delta n}{%
n_{0}},  \label{Velocityi}
\end{equation}%
and%
\begin{equation}
\frac{\delta v_{ex}}{V_{T}}=\rho k_{\perp }\frac{\Pi _{\mathrm{vex}}}{\Pi }%
\frac{\delta n}{n_{0}},~~\frac{\delta v_{ey}}{V_{T}}=i\rho k_{\perp }\frac{%
\Pi _{\mathrm{vey}}}{\Pi }\frac{\delta n}{n_{0}},~~\frac{\delta v_{ez}}{V_{T}%
}=\frac{V_{T}k_{z}}{\omega }\frac{\Pi _{\mathrm{vez}}}{\Pi }\frac{\delta n}{%
n_{0}},  \label{Velocitye}
\end{equation}%
where
\begin{eqnarray}
\Pi _{\mathrm{vix}} &=&\omega \omega _{ci}\left[ \left( 1+Q+\lambda
_{e}^{2}k^{2}\right) ^{2}\omega ^{2}-\left( 1+Q+\lambda _{i}^{2}k^{2}\right)
V_{A}^{2}k_{z}^{2}\right] ,  \notag \\
\Pi _{\mathrm{viy}} &=&\left[ Q\left( 1+Q+\lambda _{e}^{2}k^{2}\right)
\omega ^{2}-V_{A}^{2}k_{z}^{2}\right] V_{A}^{2}k^{2},  \notag \\
\Pi _{\mathrm{viz}} &=&\left( 1+Q+\lambda _{e}^{2}k^{2}\right) ^{2}\omega
^{4}-\left[ \left( 1+Q+\lambda _{i}^{2}k^{2}\right)
V_{A}^{2}k_{z}^{2}+\left( 1+Q+Q\lambda _{e}^{2}k^{2}\right) V_{A}^{2}k^{2}%
\right] \omega ^{2}+V_{A}^{4}k^{2}k_{z}^{2},  \notag \\
\Pi _{\mathrm{vex}} &=&\omega \omega _{ci}\left[ \left( 1+Q+\lambda
_{e}^{2}k^{2}\right) ^{2}\omega ^{2}-\left( 1+Q+Q\lambda
_{e}^{2}k^{2}\right) V_{A}^{2}k_{z}^{2}\right] ,  \notag \\
\Pi _{\mathrm{vey}} &=&\left[ \left( 1+Q+\lambda _{e}^{2}k^{2}\right) \omega
^{2}-QV_{A}^{2}k_{z}^{2}\right] V_{A}^{2}k^{2},  \notag \\
\Pi _{\mathrm{vez}} &=&\left( 1+Q+\lambda _{e}^{2}k^{2}\right) ^{2}\omega
^{4}-\left[ \left( 1+Q+Q\lambda _{e}^{2}k^{2}\right)
V_{A}^{2}k_{z}^{2}+\left( 1+Q+\lambda _{i}^{2}k^{2}\right) V_{A}^{2}k^{2}%
\right] \omega ^{2}+V_{A}^{2}k^{2}k_{z}^{2}.  \notag
\end{eqnarray}%
Note that we can explore the linear relation between arbitrary two variables
through the eigenfunctions (\ref{Electric-field}) and (\ref{Magnetic-field})$%
-$(\ref{Velocitye}). For example, the polarizations of electromagnetic
fields are

\begin{equation}
\frac{\delta E_{x}}{\delta E_{y}}=i\frac{\Pi _{\mathrm{Ex}}}{\Pi _{\mathrm{Ey%
}}},~~\frac{\delta E_{z}}{\delta E_{y}}=i\frac{k_{z}}{k_{\perp }}\frac{\Pi _{%
\mathrm{Ez}}}{\Pi _{\mathrm{Ey}}},~~  \label{Eq: Electric polarization}
\end{equation}%
and

\begin{equation}
\frac{\delta B_{x}}{\delta B_{y}}=-i\frac{\Pi _{\mathrm{Ey}}}{\left(
1-Q^{2}\right) V_{A}^{2}k^{2}\omega ^{2}},~~\frac{\delta B_{z}}{\delta B_{y}}%
=i\frac{k_{\perp }}{k_{z}}\frac{\Pi _{\mathrm{Ey}}}{\left( 1-Q^{2}\right)
V_{A}^{2}k^{2}\omega ^{2}}.  \label{Eq: Magnetic polarization}
\end{equation}

\subsection{Parallel Waves}

At parallel propagation, $\mathbf{k}=k_{z}\hat{\mathbf{z}}$, Eq. (\ref%
{Eq:General dispersion equation}) is written as
\begin{equation}
\left[ \left( 1+Q+\lambda _{e}^{2}k_{z}^{2}\right) ^{2}\omega ^{4}-\left(
2\left( 1+Q\right) +\left( 1+Q^{2}\right) \lambda _{i}^{2}k_{z}^{2}\right)
V_{A}^{2}k_{z}^{2}\omega ^{2}+V_{A}^{4}k_{z}^{4}\right] \left[ \left(
1+Q\right) \omega ^{2}-V_{T}^{2}k_{z}^{2}\right] =0,
\label{Eq:Parallel dispersion}
\end{equation}%
which describes the left-hand $\left( \omega _{-}\right) $ and right-hand $%
\left( \omega _{+}\right) $ circularly-polarized waves
\begin{equation}
\omega _{\pm }^{2}=V_{A}^{2}k_{z}^{2}\frac{1+Q+\left( 1+Q^{2}\right) \lambda
_{i}^{2}k_{z}^{2}/2}{\left( 1+Q+\lambda _{e}^{2}k_{z}^{2}\right) ^{2}}\left[
1\pm \sqrt{1-\left( \frac{1+Q+\lambda _{e}^{2}k_{z}^{2}}{1+Q+\left(
1+Q^{2}\right) \lambda _{i}^{2}k_{z}^{2}/2}\right) ^{2}}\right] ,
\label{ion and electron cyclotron wave}
\end{equation}%
and ion acoustic wave
\begin{equation}
\omega ^{2}=V_{T}^{2}k_{z}^{2}/\left( 1+Q\right) .  \label{ion acoustic wave}
\end{equation}%
Note that the dispersion relation (\ref{ion and electron cyclotron wave})
can be directly derived from Eqs. (\ref{Eq:Relation_x}) and (\ref%
{Eq:Relation_y}); (\ref{ion acoustic wave}) can be derived by use of Eqs. (%
\ref{Eq:Relation_z}) and (\ref{Eq:ION-EQUATION}).

The left- and right-hand waves have the perpendicular perturbations
\begin{eqnarray}
\delta E_{y} &=&\mp i\delta E_{x},  \notag \\
\delta B_{y} &=&\mp i\delta B_{x}=\frac{k_{z}}{\omega }\delta E_{x},  \notag
\\
\delta v_{iy} &=&\mp i\delta v_{ix}=-\frac{\delta E_{x}}{B_{0}\left( 1\mp
\omega /\omega _{ci}\right) },  \notag \\
\delta v_{ey} &=&\mp i\delta v_{ex}=-\frac{\delta E_{x}}{B_{0}\left( 1\pm
Q\omega /\omega _{ci}\right) },  \label{Parallel.perturbation1}
\end{eqnarray}%
whereas the ion acoustic wave has the parallel perturbations
\begin{eqnarray}
\delta E_{z} &=&-ik_{z}\frac{\gamma _{e}T_{e}-Q\gamma _{i}T_{i}}{e\left(
1+Q\right) }\frac{\delta n}{n_{0}},  \notag \\
\delta v_{iz} &=&\delta v_{ez}=\frac{\omega }{k_{z}}\frac{\delta n}{n_{0}}.
\label{Parallel.perutrbation2}
\end{eqnarray}

\subsection{Perpendicular waves}

When the wave propagates at the perpendicular direction, $\mathbf{k}%
=k_{\perp }\hat{\mathbf{x}}$, only one mode exists
\begin{equation}
\omega ^{2}=\frac{1+Q+\left( 1+Q+\lambda _{e}^{2}k_{\perp }^{2}\right) \beta
}{\left( 1+Q\right) \left( 1+Q+\lambda _{e}^{2}k_{\perp }^{2}\right) }%
V_{A}^{2}k_{\perp }^{2}.  \label{perpendicular wave}
\end{equation}%
Its polarization properties are
\begin{eqnarray}
\frac{\delta E_{x}}{B_{0}} &=&i\frac{\left( 1+Q\right) \left( Q\widetilde{T}%
_{i}-\widetilde{T}_{e}\right) \omega ^{2}-\left( \Lambda _{2}\widetilde{T}%
_{i}-Q\Lambda _{0}\widetilde{T}_{e}\right) V_{A}^{2}k_{\perp }^{2}}{\left(
1+Q+\lambda _{e}^{2}k_{\perp }^{2}\right) k_{\perp }\omega _{ci}}\frac{%
\delta n}{n_{0}},~  \notag \\
~\frac{\delta E_{y}}{B_{0}} &=&\frac{1+Q}{1+Q+\lambda _{e}^{2}k_{\perp }^{2}}%
\frac{\omega }{k_{\perp }}\frac{\delta n}{n_{0}},  \notag \\
\frac{\delta B_{z}}{B_{0}} &=&\frac{1+Q}{1+Q+\lambda _{e}^{2}k_{\perp }^{2}}%
\frac{\delta n}{n_{0}},  \notag \\
\frac{\delta v_{ix}}{V_{A}} &=&\frac{\delta v_{ex}}{V_{A}}=\frac{\omega }{%
V_{A}k_{\perp }}\frac{\delta n}{n_{0}},~~  \notag \\
\frac{\delta v_{iy}}{V_{A}} &=&-Q\frac{\delta v_{ey}}{V_{A}}=-i\frac{\sqrt{Q}%
\lambda _{e}k_{\perp }}{1+Q+\lambda _{e}^{2}k_{\perp }^{2}}\frac{\delta n}{%
n_{0}}.  \label{penpendicular.perturbation}
\end{eqnarray}

\section{Discussion}

\begin{figure}[tbp]
\includegraphics[width=16cm]{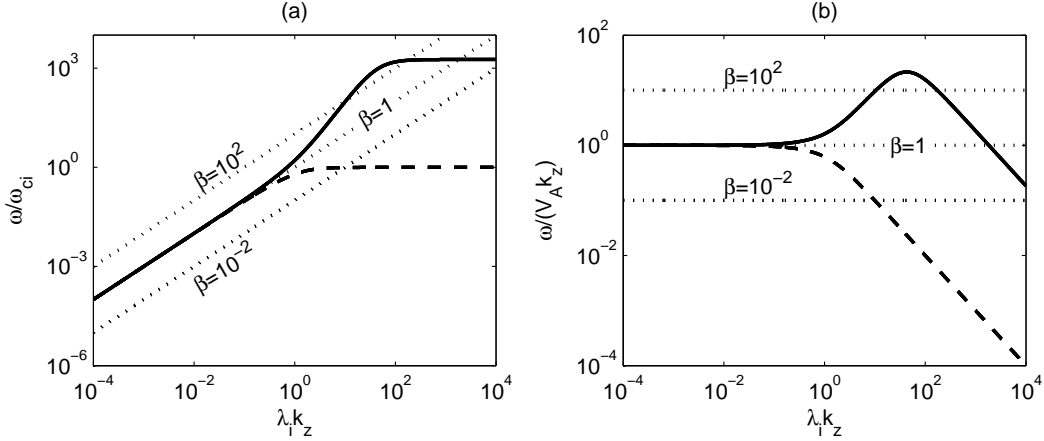}
\caption{Wave frequency and phase velocity of the parallel right-hand
circularly-polarized wave (solid lines), left-hand circularly-polarized wave
(dashed lines), and ion acoustic wave (dotted lines) in the plasmas with $%
T_{i}=T_{e}$\ but different $\protect\beta $: $\protect\beta =10^{-2}$, $1$
and $10^{2}$. }
\label{fig:1}
\end{figure}

At the parallel propagation, the ion acoustic wave can interact with the
right/left circularly-polarized waves at interaction points where their $%
\omega /\omega _{ci}$ and $\lambda _{i}k_{z}$ are equal as shown in Fig.
(1). A mode transition can occur at the interaction point. The mode
transition can happen among three oblique waves in Figs. (2)$-$(4). Also,
Figs. (2)$-$(4) include the wave electromagnetic polarizations as well as
the magnetic helicity $\sigma $ and the ion cross-helicity $\sigma _{Ci}$
\begin{equation*}
\sigma =\frac{k\left( \mathbf{A\cdot \delta B^{\ast }}\right) }{\delta B^{2}}%
=\frac{2\mathrm{cos}\theta \left( i\delta B_{x}/\delta B_{y}\right) }{%
\mathrm{cos}^{2}\theta +\left( \delta B_{x}^{2}/\delta B_{y}^{2}\right) },
\end{equation*}%
and
\begin{equation*}
\sigma _{Ci}=\frac{2\mathrm{Re}\left( \delta \mathbf{v}_{i}\cdot \delta
\mathbf{v}_{B}^{\ast }\right) }{\delta v_{i}^{2}+\delta v_{B}^{2}},
\end{equation*}%
where $\mathbf{A}$ denotes the vector potential and $\delta v_{B}\equiv
\delta B/\sqrt{\mu _{0}n_{0}m_{i}}$ is the magnetic field perturbation in
the velocity unit.

\begin{figure*}[h]
\includegraphics[width=14cm]{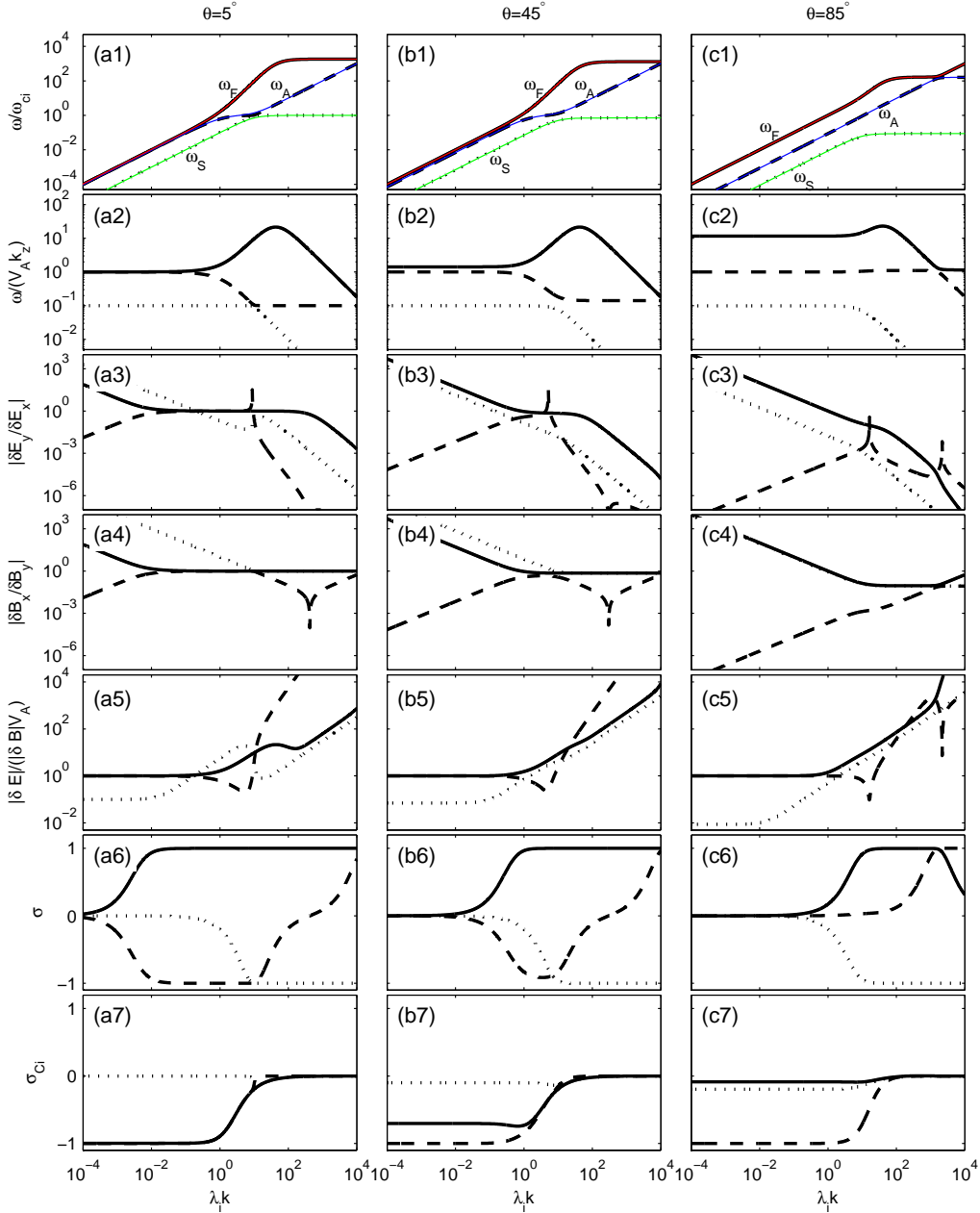}
\caption{Wave dispersion relation and polarization at three propagating
angles ($\protect\theta =5^{\circ }$, $45^{\circ }$ and $85^{\circ }$) in
the low-$\protect\beta $ plasmas with $\protect\beta =10^{-2}$ and $%
T_{e}=T_{i}$, where the solid, dashed and dotted lines denote the fast, Alfv%
\'{e}n and slow modes, respectively. Thin solid lines in Panels (a1), (b1)
and (c1) correspond to approximate dispersion relations (\protect\ref{A31})
and (\protect\ref{A33}) in the low-$\protect\beta $ ($\protect\beta \ll 1$)
limit. }
\label{fig:2}
\end{figure*}

\begin{figure*}[h]
\includegraphics[width=14cm]{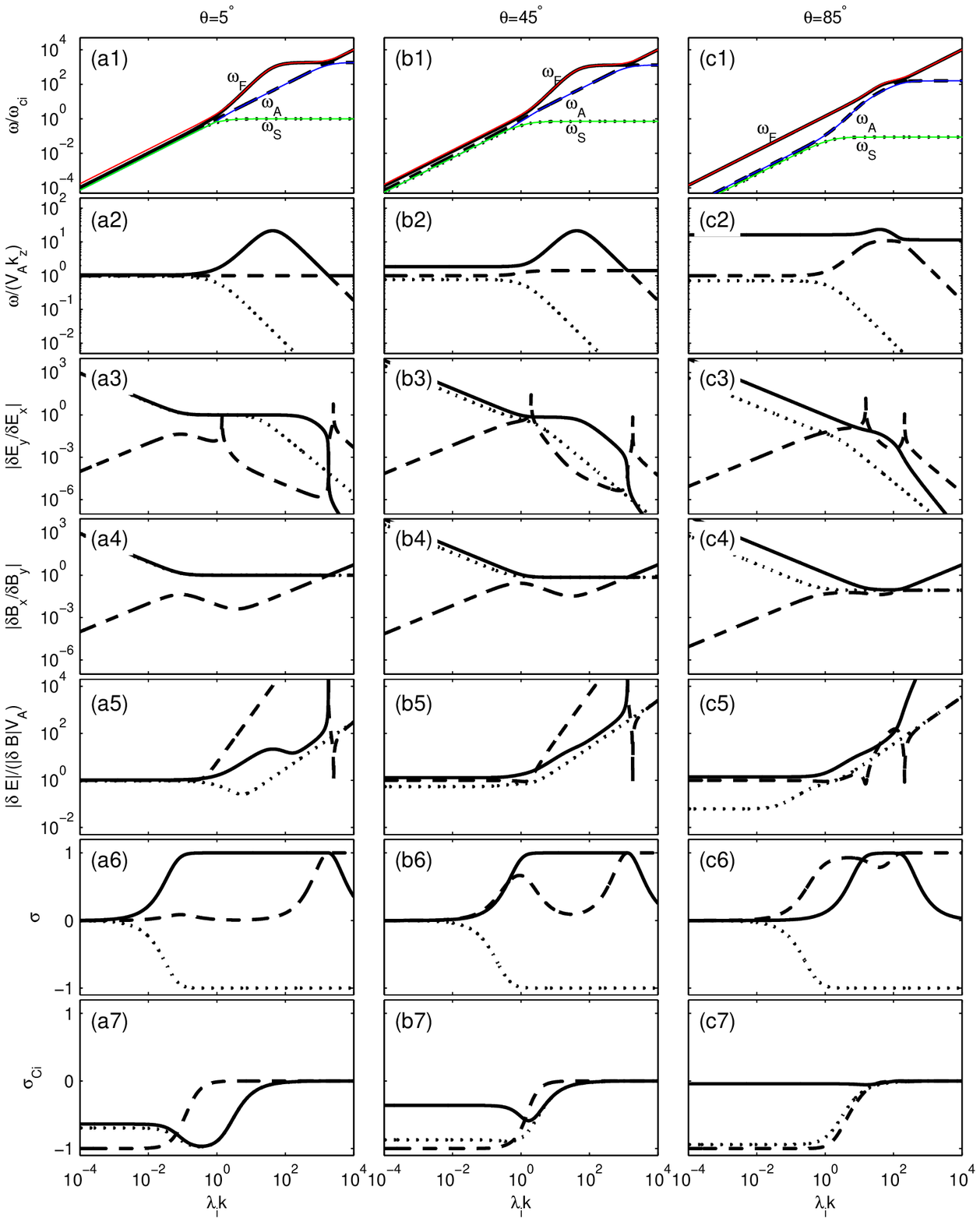}
\caption{Wave dispersion relation and polarization at three propagating
angles ($\protect\theta =5^{\circ }$, $45^{\circ }$ and $85^{\circ }$) in $%
\protect\beta =1$ plasmas with $T_{e}=T_{i}$, where the solid, dashed and
dotted lines denote the fast, Alfv\'{e}n and slow modes, respectively. Thin
solid lines in Panels (a1), (b1) and (c1) correspond to the approximate
dispersion relations (\protect\ref{A22}) and (\protect\ref{A23}).}
\label{fig:3}
\end{figure*}

\begin{figure*}[h]
\includegraphics[width=14cm]{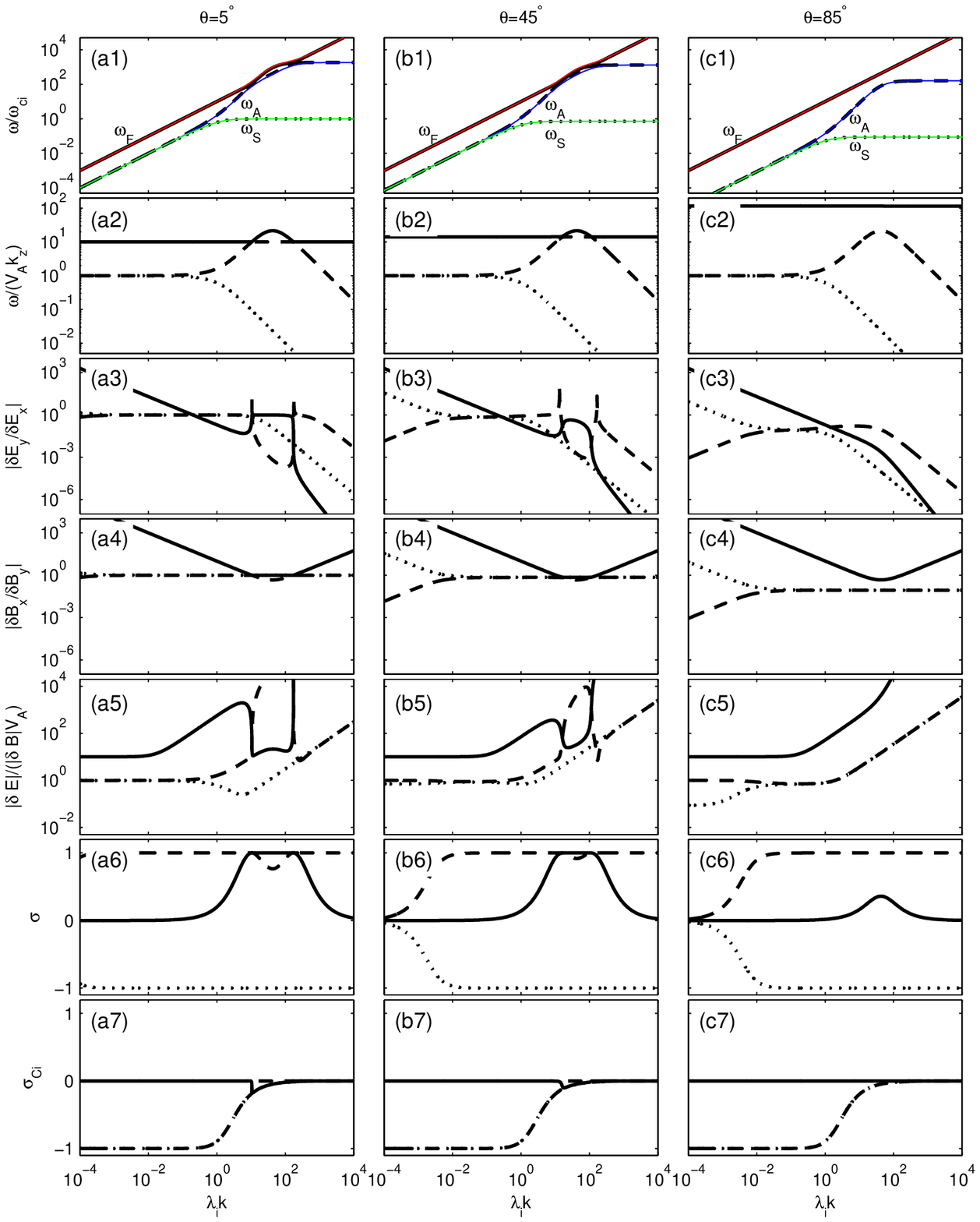}
\caption{Wave dispersion relation and polarization at three propagating
angles ($\protect\theta =5^{\circ }$, $45^{\circ }$ and $85^{\circ }$) in
the high-$\protect\beta $ plasmas with $\protect\beta =10^{2}$ and $%
T_{e}=T_{i}$, where the solid, dashed and dotted lines denote the fast, Alfv%
\'{e}n and slow modes, respectively. Thin solid lines in Panels (a1), (b1)
and (c1) represent the analytical dispersion relations (\protect\ref{A22})
and (\protect\ref{A23}) corresponding to the high-$\protect\beta $ ($\protect%
\beta \gg 1$) limit.}
\label{fig:4}
\end{figure*}

Fig. (2) presents the dispersion relations and polarizations of the three
oblique waves at different angles in the low-$\beta $ plasmas where $\beta
=10^{-2}$ and $T_{i}=T_{e}$. It shows that approximate dispersion relations (%
\ref{A31}) and (\ref{A33}) can describe the exact one (\ref{Eq:Roots}) well.

The fast mode corresponds to the fast magnetosonic wave as $\omega \ll
\omega _{ci}$ and the whistler wave as $\omega _{ci}\ll \omega \ll
\left\vert \omega _{ce}\right\vert \cos \theta $. At the electron cyclotron
frequency $\omega =\left\vert \omega _{ce}\right\vert \cos \theta $, the
fast mode is the electron cyclotron wave. Furthermore, the electron
cyclotron wave can change to a (quasi-) electroacoustic wave extending to
higher frequency $\omega >\left\vert \omega _{ce}\right\vert \cos \theta $
\cite{fk69}. It is interesting to see that the fast magnetosonic wave has $%
\left( \left\vert \delta E_{y}/\delta E_{x}\right\vert ,\hspace{0in}%
\left\vert \delta B_{x}/\delta B_{y}\right\vert \right) >1$ as $\lambda
_{i}k<10^{-2}$, and $\left( \left\vert \delta E_{y}/\delta E_{x}\right\vert ,%
\hspace{0in}\left\vert \delta B_{x}/\delta B_{y}\right\vert \right) \simeq 1$
as $10^{-2}<\lambda _{i}k\ll 1$ at the near-parallel propagation. $%
\left\vert \delta E/\delta B\right\vert \simeq V_{A}$ and $\sigma \simeq 0$
for the fast magnetosonic wave; $\left\vert \delta E/\delta B\right\vert
>V_{A}$ and $\sigma \simeq 1$ for the whistler and electron cyclotron waves.
In addition, $\sigma _{{Ci}}\simeq -\cos \theta $ for the fast magnetosonic
wave \cite{kr94}, while $\sigma _{{Ci}}\simeq 0$ for the whistler and
(quasi-) electroacoustic waves.

The Alfv\'{e}n mode is the shear Alfv\'{e}n wave at $\omega \ll \omega _{ci}$
and the (quasi-) electroacoustic wave at $\omega _{ci}<\omega <\left\vert
\omega _{ce}\right\vert \cos \theta $ until a transition into the electron
cyclotron wave at $\omega =\left\vert \omega _{ce}\right\vert \cos \theta $.
At near-parallel $\theta =5^{\circ }$ and oblique $\theta =45^{\circ }$
cases, the phase velocity of the (quasi-) electroacoustic wave is about the
sound speed. It is the Alfv\'{e}n speed at the high oblique angle,
therefore, Ref. \cite{zhao2014} called the high oblique mode at $\omega
_{ci}<\omega <\left\vert \omega _{ce}\right\vert \cos \theta $ as the shear
Alfv\'{e}n wave. Note that an ion cyclotron wave $\omega \simeq \omega _{ci}$
arises at near-parallel propagation (Panel (a1)).\ Electromagnetic
polarizations are $\left\vert \delta E_{x}\right\vert \gg \left\vert \delta
E_{y}\right\vert $, $\left\vert \delta B_{y}\right\vert \gg \left\vert
\delta B_{x}\right\vert $ and $\left\vert \delta E/\delta B\right\vert =V_{A}
$ at $\omega \ll \omega _{ci}$; at $\omega _{ci}<\omega <\left\vert \omega
_{ce}\right\vert \cos \theta $, $\left\vert \delta E/\delta B\right\vert $
becomes much larger than $V_{A}$. At $\theta =5^{\circ }$ and $45^{\circ }$,
the magnetic-helicity $\sigma $ decreases firstly from $\sigma =0$ to $%
\sigma \simeq -1$ at $\omega \ll \omega _{ci}$, and then increases with
increasing $\lambda _{i}k$ at $\omega _{ci}<\omega <\left\vert \omega
_{ce}\right\vert \cos \theta $; at $\theta =85^{\circ }$, $\sigma =0$ is
nearly unchanged at $\omega \ll \omega _{ci}$, and it becomes increasing at $%
\omega >\omega _{ci}$ until reaching $\sigma =1$ corresponding to the
electron cyclotron wave\textbf{.} Besides, the ion cross helicity $\sigma _{{%
Ci}}$ depends on the wave scale, e.g., $\sigma _{Ci}=-1$ as $\lambda
_{i}k\ll 1$ and $\sigma _{{Ci}}=0$ as $\lambda _{i}k\gg 1$.

The slow mode corresponds to the slow magnetosonic wave at $\omega \ll
\omega _{ci}$, where $\left\vert \delta E_{x}\right\vert <\left\vert \delta
E_{y}\right\vert $, $\left\vert \delta B_{x}\right\vert >\left\vert \delta
B_{y}\right\vert $, $\left\vert \delta E/\delta B\right\vert \sim V_{T}$ and
$\sigma =0$. It turns to the ion cyclotron wave at $\omega =\omega _{ci}\cos
\theta $ \cite{bs03}, where $\left\vert \delta E_{x}\right\vert \gg
\left\vert \delta E_{y}\right\vert $, $\left\vert \delta B_{x}\right\vert
\sim \left\vert \delta B_{y}\right\vert $, $\left\vert \delta E/\delta
B\right\vert >V_{A}$ and $\sigma =-1$.\ At $\theta =5^{\circ }$, the
electric polarization $\left\vert \delta E_{y}/\delta E_{x}\right\vert $\
has an increment at the transition position where the slow magnetosonic wave
changes to the ion cyclotron wave; however, there is no such increment at $%
\theta =45^{\circ }$\ and $\theta =85^{\circ }$\ cases.

Figs. (3) and (4) present the dispersion relations and polarizations in $%
\beta =1$ and high-$\beta $ $\left( \beta =10^{2}\right) $ plasmas. Here the
Alfv\'{e}n mode interacts with the fast mode only. Although the validity
condition for approximate dispersion relations (\ref{A22}) and (\ref{A23})
is $k_{\perp }^{2}/k_{z}^{2}>>1$ or $\beta \gg 1$, these expressions can
describe the wave dispersion relations at $\beta =1$ as shown in Fig. (3).

Several mode properties in Fig. (3) are obviously different from that in the
low-$\beta $ plasmas (Fig. (2)). For example, to the near-parallel waves at $%
0.1\omega _{ci}<\omega <\omega _{ci}$, two circularly polarized $\left(
\left\vert \delta E_{x}\right\vert \simeq \left\vert \delta E_{y}\right\vert
\right) $ modes are the fast and slow modes in $\beta =1$ plasmas but the
fast and Alfv\'{e}n modes in the low-$\beta $ plasmas. Here the fast (slow)
mode exhibits the right-hand (left-hand) electric polarization and positive
(negative) helicity. It also finds $\sigma \geq 0$ for the Alfv\'{e}n mode
in $\beta =1$ plasmas. Moreover, when the wave tends to more oblique
propagation, the ion cross-helicity of the slow magnetosonic wave $\sigma _{{%
Ci}}\rightarrow -1$.

Fig. (4) shows that the Alfv\'{e}n and slow modes are circularly polarized $%
\left( \left\vert \delta E_{x}\right\vert \simeq \left\vert \delta
E_{y}\right\vert \right) $ waves at $\omega <\omega _{ci}$ in the high-$%
\beta $ plasmas, where the Alfv\'{e}n (slow) mode exhibits the right-hand
(left-hand) electric polarization and positive (negative) helicity. These
two modes also have the same ion cross-helicity distribution. Note that
three modes have no interaction point at the high oblique propagation as
shown in Panel (c1).

\begin{figure*}[h]
\includegraphics[width=14cm]{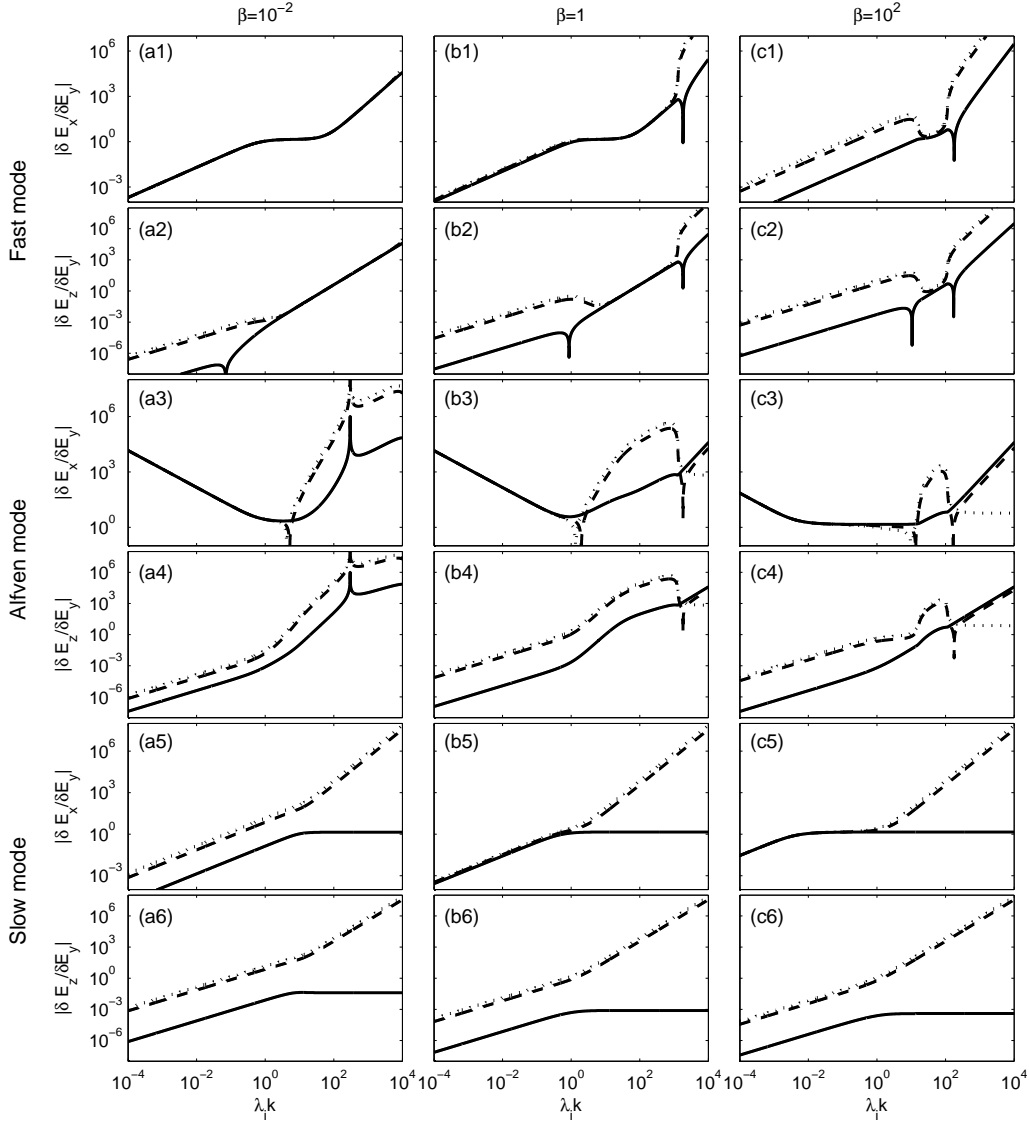}
\caption{The sensitivity of the electric polarizations of the three modes on
the temperature ratio $T_{e}/T_{i}$, where the waves propagate at $\protect%
\theta =45^{\circ }$ in the plasmas with different $\protect\beta $
environments: low-$\protect\beta $ ($\protect\beta =10^{-2}$), $\protect%
\beta =1$, and high-$\protect\beta $ ($\protect\beta =10^{2}$). The solid,
dashed and dotted lines represent the cold electron $T_{e}/T_{i}=0$, equal
ion and electron temperature $T_{e}/T_{i}=1$, and cold ion $%
T_{e}/T_{i}=\infty $ cases, respectively.}
\label{fig:5}
\end{figure*}

\begin{figure*}[h]
\includegraphics[width=14cm]{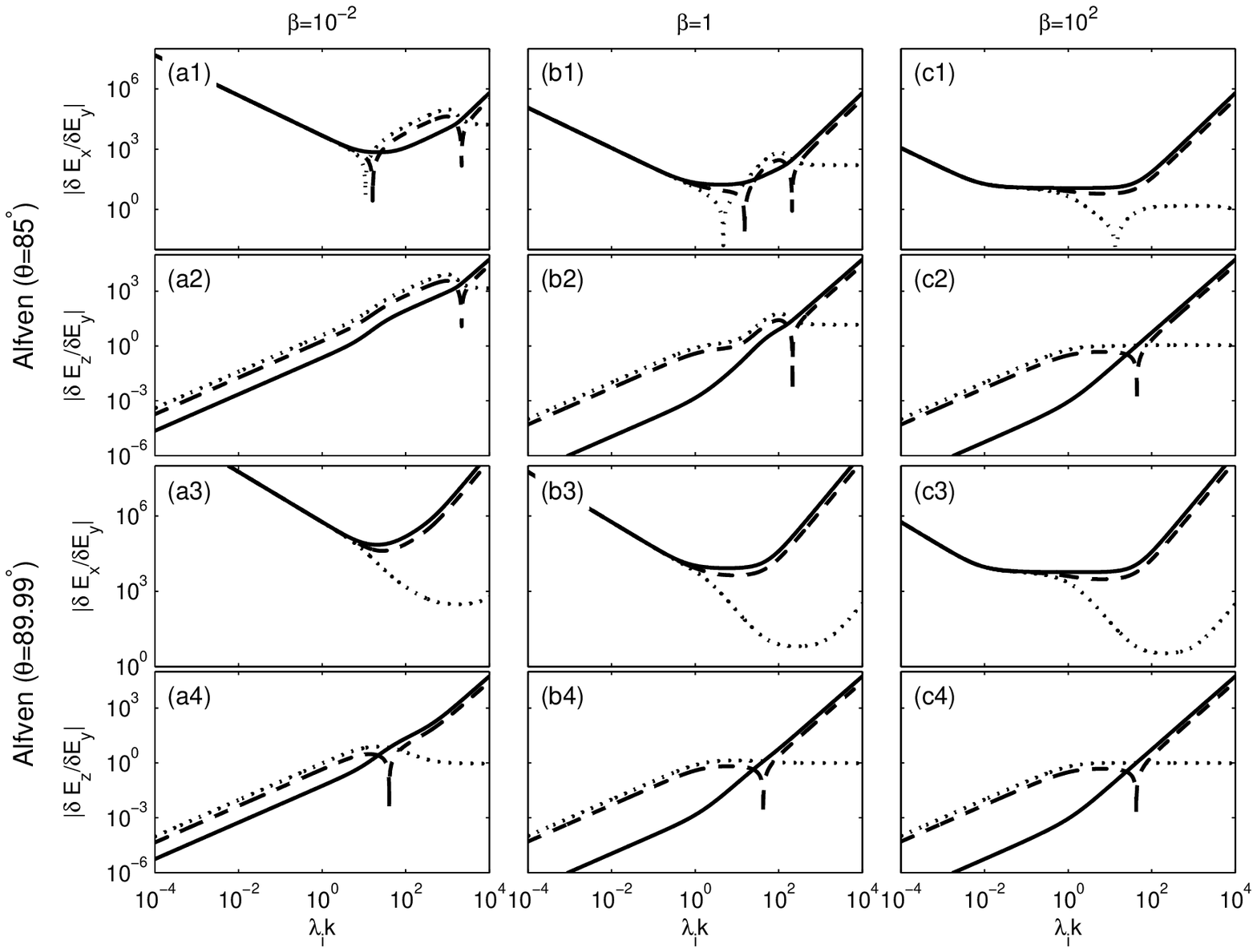}
\caption{The dependence of the electric polarizations of high oblique Alfv\'{%
e}n waves on the temperature ratio $T_{e}/T_{i}$, where two angles $\protect%
\theta =85^{\circ }$ and $89.99^{\circ }$ are considered. The solid, dashed
and dotted lines represent the cold electron $T_{e}/T_{i}=0$, equal ion and
electron temperature $T_{e}/T_{i}=1$, and cold ion $T_{e}/T_{i}=\infty $
cases, respectively.}
\label{fig:6}
\end{figure*}

It needs to note that the electric field polarizations also strongly depend
on the ratio of the electron to ion temperature $T_{e}/T_{i}$\ (Eq. (\ref%
{Eq: Electric polarization})). The electric field polarizations with
different $T_{e}/T_{i}$\ are presented in Fig. (5), where $\theta =45^{\circ
}$\ and $T_{e}/T_{i}=0$, $1$, and $\infty $. It shows that the parallel
polarization $\left\vert \delta E_{z}/\delta E_{y}\right\vert $\ decreases
obviously with decreasing $T_{e}/T_{i}$. The transverse polarization $%
\left\vert \delta E_{x}/\delta E_{y}\right\vert $\ is slightly affected by $%
T_{e}/T_{i}$\ for the long-wavelength $\lambda _{e}k\ll 1$ waves, but not
for the long-wavelength fast mode in the high-$\beta $ plasmas or for the
long-wavelength slow mode in the low-$\beta $\ plasmas. To understand
qualitatively above results, the complete expressions Eq. (\ref{Eq: Electric
polarization}) can reduce to

\begin{eqnarray}
\frac{\delta E_{x}}{i\delta E_{y}} &=&-\frac{\widetilde{T}_{i}\omega
^{2}V_{A}^{2}k^{2}+\widetilde{T}_{e}\left[ \omega ^{2}\left( \omega
^{2}-V_{A}^{2}k_{z}^{2}\right) +V_{A}^{4}k^{2}k_{z}^{2}\right] }{\left(
\omega ^{2}-V_{A}^{2}k_{z}^{2}\right) \omega \omega _{ci}},  \notag \\
\frac{\delta E_{z}}{i\delta E_{y}} &=&-\frac{k_{z}}{k_{\perp }}\frac{\left(
\omega ^{2}+V_{A}^{2}k^{2}\right) }{\omega \omega _{ci}}\widetilde{T}_{e},
\end{eqnarray}%
where the long-wavelength $\left( \lambda _{e}k\ll 1\right) $\ and very
low-frequency $\left( \Lambda _{1,2}\simeq 1\right) $\ conditions are used,
and the smaller terms of the order of $Q$\ are neglected. Since $\widetilde{T%
}_{i}=1/\left( 1+\gamma _{e}T_{e}/\gamma _{i}T_{i}\right) $\ and $\widetilde{%
T}_{e}=\left( 1-\widetilde{T}_{i}\right) $, $\left\vert \delta E_{z}/\left(
i\delta E_{y}\right) \right\vert $\ decreases with decreasing $T_{e}/T_{i}$.
When $\omega \sim V_{A}k_{z}$, $\delta E_{x}/\left( i\delta E_{y}\right)
\sim -V_{A}^{2}k^{2}\lambda _{i}k_{z}/\left( \omega
^{2}-V_{A}^{2}k_{z}^{2}\right) $ is independent on $T_{e}/T_{i}$. We can
also find $\delta E_{x}/\left( i\delta E_{y}\right) \sim -\widetilde{T}%
_{e}\omega /\omega _{ci}$ corresponding to the fast wave $\omega _{F}\sim
V_{T}k$ in $\beta \gg 1$ plasmas\ and $\delta E_{x}/\left( i\delta
E_{y}\right) \sim \widetilde{T}_{e}V_{A}^{2}k^{2}/\omega \omega _{ci}$\
corresponding to the slow wave $\omega _{S}\sim V_{T}k_{z}$\ in $\beta \ll 1$%
\ plasmas, which indicate $\left\vert \delta E_{x}/\left( i\delta
E_{y}\right) \right\vert $\ decreasing with smaller $T_{e}/T_{i}$. Besides,
in Fig. (5) the electric polarizations $\left\vert \delta E_{x}/\delta
E_{y}\right\vert $\ and $\left\vert \delta E_{z}/\delta E_{y}\right\vert $\
in $T_{e}/T_{i}\neq 0$ plasmas increase continuously as the slow
magnetosonic wave changes to the ion cyclotron wave, while both
polarizations are nearly unchanged in $T_{e}/T_{i}=0$ plasmas. Note that the
main characters in Fig. (5) still appear in the electric polarization
distributions of the oblique waves with $\theta \neq 45^{\circ }$.

When the phase relation of the electric polarizations changes, a peak or a
valley can occur in $\left\vert \delta E_{x}/\delta E_{y}\right\vert $\ and $%
\left\vert \delta E_{z}/\delta E_{y}\right\vert $ distributions. For the
fast mode in Fig. (5), the phase relation of $\delta E_{z}/\delta E_{y}$\
changes in $T_{e}/T_{i}=0$\ plasmas, while it is unchanged in $%
T_{e}/T_{i}\neq 0$\ plasmas. For the Alfv\'{e}n mode, two transition points
arise in the phase relation of $\delta E_{x}/\delta E_{y}$ in $%
T_{e}/T_{i}\neq 0$\ plasmas; however, in the cold electron $\left(
T_{e}/T_{i}=0\right) $\ plasmas, the transition at the smaller $\lambda
_{i}k $\ disappears in $\beta \ll 1$\ plasmas, or two transitions are both
missing in $\beta \geq 1$\ plasmas.

For near-perpendicular Alfv\'{e}n wave, Fig. (6) shows that the parallel
polarization $\left\vert \delta E_{z}/\delta E_{y}\right\vert $\ increases
(decreases) with $T_{e}/T_{i}$\ as $\lambda _{i}k<10^{2}$\ $\left( \lambda
_{i}k>10^{2}\right) $. At $\theta =89.99^{\circ }$, the transverse
polarization $\left\vert \delta E_{x}/\delta E_{y}\right\vert $\ decreases
with increasing $T_{e}/T_{i}$ for the kinetic-scale Alfv\'{e}n waves $\left(
\lambda _{i}k>1\right) $. These results indicate the important role of the
electron temperature $T_{e}$\ on the kinetic-scale Alfv\'{e}n waves \cite%
{le99,ya14}. Moreover, there is no transition of the phase relation of $%
\delta E_{x}/\delta E_{y}$\ at $\theta =89.99^{\circ }$\ case. The reason is
that the wave frequency $\omega $ is smaller than the ion cyclotron
frequency $\omega _{ci}$ at $\theta =89.99^{\circ }$, which cannot satisfy
the frequency condition $\omega >\omega _{ci}$ for the changing of the phase
relation of transverse electric polarization\ \cite{zhao2014}.

\section{Summary}

In this study ions and electrons are treated separately in comparison with
one fluid element $\left( U=\sum m_{\alpha }v_{\alpha }/\sum m_{\alpha
}\right) $ method adopted in previous studies \cite{st63,be12}. This method
is helpful to obtain the linear eigenfunctions including the ion and
electron velocities as well as the ion and electron cross-helicities.

It found that the fast and Alfv\'{e}n modes are nearly linearly polarized at
the very low-frequency $\omega \ll 10^{-2}\omega _{ci}$, and circularly
polarized at $10^{-2}\omega _{ci}<\omega <\omega _{ci}$ at the near-parallel
propagation in the low-$\beta $ plasmas. Two circularly polarized modes
become the fast and slow modes in a narrow frequency regime $0.1\omega
_{ci}<\omega <\omega _{ci}$ in $\beta =1$ plasmas; they are the Alfv\'{e}n
and slow modes in $\beta =10^{2}$ plasmas. To the ion cross-helicity $\sigma
_{{Ci}}$ of the long-wavelength slow mode, $\sigma _{{Ci}}\simeq 0$ in the
low-$\beta $ plasmas, $\sigma _{{Ci}}\rightarrow -1$ as $\theta \rightarrow
90^{\circ }$ in $\beta =1$ plasmas, and $\sigma _{{Ci}}\simeq -1$ in the
high-$\beta $ plasmas. It also found that the\ negative magnetic-helicity $%
\sigma $\ of the Alfv\'{e}n mode can occur at the small or moderate angles
in the low-$\beta $\ plasmas, while $\sigma \geq 0$ arises always at the
high oblique angle in the low-$\beta $\ plasmas or at the general angle in $%
\beta \geq 1$\ plasmas.

Our results exhibited the sensitivity of the electric polarizations on the
temperature ratio $T_{e}/T_{i}$. The parallel polarization $\left\vert
\delta E_{z}/\delta E_{y}\right\vert $\ decreases with $T_{e}/T_{i}$\ as $%
\lambda _{i}k<1$. The transverse polarizations $\left\vert \delta
E_{x}/\delta E_{y}\right\vert $\ also decreases with $T_{e}/T_{i}$\ for the
long-wavelength fast mode in the high-$\beta $ plasmas, or for the
long-wavelength slow mode in the low-$\beta $\ plasmas, while $\left\vert
\delta E_{x}/\delta E_{y}\right\vert $\ at other long-wavelength cases are
slightly affected by $T_{e}/T_{i}$. Furthermore, the phase relation of $%
\delta E_{x}/\delta E_{y}$ of the Alfv\'{e}n mode will change in $%
T_{e}/T_{i}\neq 0$\ plasmas, but this change can disappear in the cold
electron $T_{e}/T_{i}=0$ plasmas. For the fast mode, the phase relation of $%
\delta E_{z}/\delta E_{y}$ changes in $T_{e}/T_{i}=0$\ plasmas, while the
unchanged phase relation arises in $T_{e}/T_{i}\neq 0$\ plasmas.

We have also presented the approximate dispersion relations in the
near-perpendicular propagation, low-$\beta $, and high-$\beta $ limits.
These approximations can describe nicely\ the exact dispersion relations of
the three modes given by Eq. (\ref{Eq:Roots}). It notes that the condition
of $V_{A}^{2}/c^{2}\ll 1$ used in the study leads to the neglecting of the
displacement current. However, this assumption is broken near the wave
cutoff position which results in the validity condition of $\omega \ll
\omega _{ci}\left( 1+c^{2}/V_{A}^{2}\right) $ \cite{be12}. Also, the
displacement current may be important in producing the parallel electric
field of the low-frequency Alfv\'{e}n mode \cite{sl06}. Therefore, a
comprehensive study including the effect of the displacement current is
needed.

Lastly, two-fluid model neglects the kinetic wave-particle interaction
effects, such as Landau damping and ion (electron) cyclotron resonance
damping, which can only be captured by the kinetic model. These kinetic
effects can strongly affect the wave dispersion relation and polarization
properties. For example, the wave dispersion relation of the kinetic Alfv%
\'{e}n wave is depressed at ion scales in the high-$\beta $\ plasmas where
there can be the heavy Landau damping \cite{ho06}. The two models also
result in different phase relation between two electric components. However,
since it is difficult to identify clearly all modes from the full kinetic
theory, the two-fluid theory can be a useful guide to discard the modes in
the kinetic theory. Our complete expressions can be conveniently used to
compare with the results of the kinetic model.

\appendix

\section{Approximate dispersion relations in different limits}

\subsection{Near-perpendicular propagation limit}

At the near-perpendicular propagation limit $k_{z}^{2}/k_{\perp }^{2}\ll 1$,
the cubic equation (\ref{Eq:General dispersion equation}) for $\omega ^{2}$
can reduce to a quadratic equation

\begin{equation}
A_{\mathrm{obli}}\omega ^{4}-B_{\mathrm{obli}}\omega ^{2}+C_{\mathrm{obli}%
}=0,  \label{A21}
\end{equation}%
where
\begin{eqnarray}
A_{\mathrm{obli}} &=&1+\lambda _{e}^{2}k^{2}+\lambda
_{i}^{2}k_{z}^{2}+\left( 1+\lambda _{e}^{2}k^{2}\right) ^{2}\beta
+k_{z}^{2}/k^{2},  \notag \\
B_{\mathrm{obli}} &=&\left( 1+2\beta +\rho ^{2}k^{2}\right)
V_{A}^{2}k_{z}^{2},  \notag \\
C_{\mathrm{obli}} &=&\beta V_{A}^{4}k_{z}^{4},  \notag
\end{eqnarray}%
which contains the dispersion relation of the Alfv\'{e}n mode $\left( \omega
_{A}=\omega _{+}\right) $ and slow mode $\left( \omega _{S}=\omega
_{-}\right) $,

\begin{eqnarray}
\omega _{\pm }^{2} &=&V_{A}^{2}k_{z}^{2}\frac{1+2\beta +\rho ^{2}k^{2}}{%
2\left( 1+\lambda _{e}^{2}k^{2}+\lambda _{i}^{2}k_{z}^{2}+\left( 1+\lambda
_{e}^{2}k^{2}\right) ^{2}\beta +k_{z}^{2}/k^{2}\right) }  \notag \\
&&\left[ 1\pm \sqrt{1-4\beta \frac{1+\lambda _{e}^{2}k^{2}+\lambda
_{i}^{2}k_{z}^{2}+\left( 1+\lambda _{e}^{2}k^{2}\right) ^{2}\beta
+k_{z}^{2}/k^{2}}{\left( 1+2\beta +\rho ^{2}k^{2}\right) ^{2}}}\right]
\label{A22}
\end{eqnarray}%
On the other hand, the fast mode can be obtained by first two terms in Eq. (%
\ref{Eq:General dispersion equation}),

\begin{equation}
\omega _{F}=V_{A}k\frac{\sqrt{1+\lambda _{e}^{2}k^{2}+\lambda
_{i}^{2}k_{z}^{2}+\left( 1+\lambda _{e}^{2}k^{2}\right) ^{2}\beta
+k_{z}^{2}/k^{2}}}{1+\lambda _{e}^{2}k^{2}}.  \label{A23}
\end{equation}

\subsection{Low-$\protect\beta $ $\left( \protect\beta \ll 1\right) $ limit}

In low-$\beta $ plasmas with $\beta \ll 1$, the frequency of the slow mode $%
\omega _{S}\sim V_{T}k_{z}$ is much smaller than that of the fast mode $%
\omega _{F}\sim V_{A}k$ and Alfv\'{e}n mode $\omega _{A}\sim V_{A}k_{z}$. So
that the last two terms in Eq. (\ref{Eq:General dispersion equation})
control the dispersion relation of the slow mode

\begin{equation}
\omega _{S}=\frac{V_{T}k_{z}}{\sqrt{1+\rho ^{2}k^{2}}}.  \label{A31}
\end{equation}%
On the other hand, the fast and Alfv\'{e}n modes are controlled by following
quadratic equation

\begin{equation}
A_{\mathrm{low}}\omega ^{4}-B_{\mathrm{low}}\omega ^{2}+C_{\mathrm{low}}=0,
\label{A32}
\end{equation}%
where
\begin{eqnarray}
A_{\mathrm{low}} &=&\left( 1+\lambda _{e}^{2}k^{2}\right) ^{2},  \notag \\
B_{\mathrm{low}} &=&\left( 1+\lambda _{e}^{2}k^{2}+\left( 1+\lambda
_{e}^{2}k^{2}\right) ^{2}\beta +\lambda
_{i}^{2}k_{z}^{2}+k_{z}^{2}/k^{2}\right) V_{A}^{2}k^{2},  \notag \\
C_{\mathrm{low}} &=&\left( 1+\rho ^{2}k^{2}\right) V_{A}^{4}k^{2}k_{z}^{2},
\notag
\end{eqnarray}%
which yields the dispersion relations for the fast mode $\left( \omega
_{F}=\omega _{+}\right) $ and Alfv\'{e}n mode $\left( \omega _{A}=\omega
_{-}\right) $

\begin{eqnarray}
\omega _{\pm }^{2} &=&V_{A}^{2}k^{2}\frac{1+\lambda _{e}^{2}k^{2}+\left(
1+\lambda _{e}^{2}k^{2}\right) ^{2}\beta +\lambda
_{i}^{2}k_{z}^{2}+k_{z}^{2}/k^{2}}{2\left( 1+\lambda _{e}^{2}k^{2}\right)
^{2}}  \notag \\
&&\left[ 1\pm \sqrt{1-\frac{4\left( 1+\lambda _{e}^{2}k^{2}\right)
^{2}\left( 1+\rho ^{2}k^{2}\right) }{\left( 1+\lambda _{e}^{2}k^{2}+\left(
1+\lambda _{e}^{2}k^{2}\right) ^{2}\beta +\lambda
_{i}^{2}k_{z}^{2}+k_{z}^{2}/k^{2}\right) ^{2}}\frac{k_{z}^{2}}{k^{2}}}\right]
.  \label{A33}
\end{eqnarray}

\subsection{High-$\protect\beta $ $\left( \protect\beta \gg 1\right) $ limit}

In high-$\beta $ plasmas with $\beta \gg 1$, the frequency of the fast mode $%
\omega _{F}\sim V_{T}k$\ is much larger than that of the Alfv\'{e}n and slow
modes $\omega _{A,S}\sim V_{A}k_{z}$. The fast mode are dominant by the
first two terms in Eq. (\ref{Eq:General dispersion equation}), whereas the
Alfv\'{e}n and slow modes are dominant by the quadratic equation shown in
Eq. (\ref{A21}). Therefore, the wave dispersion relations of the three modes
are the same as that given by Eqs. (\ref{A22}) and (\ref{A23}).


%

\begin{acknowledgments}
The author thanks Prof. M. Y. Yu for discussing and improving the paper. The
author also thanks the anonymous referee for constructive comments and
suggestions that improve the quality of the paper. This work was supported
by the NNSFC 11303099, the NSF of Jiangsu Province (BK2012495), and the Key
Laboratory of Solar Activity at CAS NAO (LSA201304).
\end{acknowledgments}


\clearpage

%
%

\end{document}